Discordance between genomic divergence and phenotypic variation in a rapidly evolving avian genus (*Motacilla*)

Rebecca B. Harris[1,2*], Per Alström[3,4,5], Anders Ödeen[3], and Adam D. Leaché[1,2]

[1] Department of Biology, University of Washington, Seattle, WA 98195
[2] Burke Museum of Natural History and Culture, Seattle, WA 98195
[3] Department of Animal Ecology, Evolutionary Biology Centre, Uppsala University, Uppsala, Sweden
[4] Swedish Species Information Centre, Swedish University of Agricultural Sciences, Box 7007, Uppsala SE-750 07, Sweden
[5] Key Laboratory of Zoological Systematics and Evolution, Institute of Zoology, Chinese Academy of Sciences, Beijing 100101, China

* Corresponding author: rbharris@uw.edu

*We dedicate this work to Anders Ödeen, a dear friend, colleague, and pioneer in wagtail genetics, who passed away during the preparation of this manuscript.*

**Abstract**

Generally, genotypes and phenotypes are expected to be spatially congruent; however, in widespread species complexes with few barriers to dispersal, multiple contact zones, and limited reproductive isolation, discordance between phenotypes and phylogeographic groups is more probable. Wagtails (Aves: *Motacilla*) are a genus of birds with striking plumage pattern variation across Eurasia. Up to 13 subspecies are recognized within a single species, yet previous studies using mitochondrial DNA have supported phylogeographic groups that are inconsistent with subspecies plumage characteristics. In this study, we investigate the link between phenotypes and genotype by comparing populations thought to be at different stages along the speciation continuum. We take a phylogeographic approach by estimating population structure, testing for isolation by distance, conducting demographic modeling, and estimating the first time-calibrated species tree for the genus. Our study provides strong evidence for species-level patterns of differentiation in wagtails, however population-level differentiation is less pronounced. We find evidence that three of four widespread Eurasian species exhibit an east-west divide that contradicts both subspecies taxonomy and phenotypic variation. Both the geographic location of this divide and time estimates from demographic models are overlapping in two sympatric species, indicating that coincident Pleistocene events shaped their histories.

**Introduction**

Species complexes are often characterized by high frequencies of hybridization and poorly developed isolation barriers, despite being structured geographically [1]. This can lead to incongruence between genotypes and phenotypes [2]. Lack of overall genetic differentiation in taxa with distinct phenotypic differences is likely due to either (1) recent divergence, with strong selection on phenotype, or (2) large-scale introgression, except on pre-existing adaptive genetic differences. Non-adaptive demographic processes will have a universal and random effect on the genome [3], whereas selection may cause differentiation at few adaptive loci [4]. Therefore, early in the speciation process, patterns of genetic divergence may be inconsistent with the history of population divergence [5].

In birds, male plumage differences among closely related taxa are often believed to be the result of sexual selection and to play an important role in reproductive isolation [1]. Plumage differences can evolve rapidly [6–8], when populations are geographically structured, may result from spatial variation in selection regimes [1]. Recent studies have demonstrated that a small number of genes can cause dramatic plumage differences despite limited genetic differentiation throughout the remainder of the genome [9–12]. Widespread avian species complexes with marked plumage variation are useful systems for investigating heterogeneity in genomic divergence and geographic variation in phenotypes because they offer comparisons between presumably easy to identify populations at different stages of the speciation continuum.

One bird system that is particularly well suited for such studies is the passerine genus *Motacilla*. *Motacilla* consists of 12 species distributed throughout the Old World [13,14] that have earned the common name wagtail due to their propensity to pump their long tails up and down. The four sympatric, migratory wagtail species (*M. alba, M. cinerea, M.citreola*, and *M. flava*) widely distributed across the Palearctic during the breeding season represent a striking contrast in spatial variation in male plumage (cf. Fig. 1). Currently, subspecies are defined by differences in both color and pattern of head plumage in the *M. flava* complex (13 subspecies) and by head, back, and wing-covert plumage in *M. alba* (9 subspecies). On the basis of both genetic and plumage data, many of these subspecies have been treated as separate species (reviewed in [13]). Plumage differences are thought to have evolved rapidly and in conflict with phylogeographic structure [15–19]. In contrast, the other two Palearctic breeding species, *M. cinerea* and *M. citreola*, lack this extreme plumage variation.

Previous phylogenetic and phylogeographic studies of *Motacilla* report mitochondrial relationships incongruent with both taxonomy [15,17–19] and nuclear relationships [16,20,21], with suggestions that mitochondrial DNA (mtDNA) poorly reflects the true phylogeny. Several of these studies focused on aspects of wagtails' plumage diversity, some proposing cases of remarkable parallel plumage evolution [13,20,21] and others implicating the role of selection in rapid plumage evolution [15]. The *M. alba* species complex is thought to belong to a "black-and-white" plumage clade, along with three monotypic species with rather restricted allopatric distributions in the Indian subcontinent (*M. maderaspatensis*), Cambodia (*M. samveasnae*), and Japan (*M. grandis*). The *M. flava* species complex has repeatedly been found paraphyletic with *M. citreola.* A more recent phylogenetic exploration of the genus found the São Tomé endemic *Amaurocichla bocagii* nested within *Motacilla*, and proposed its inclusion within the genus [22]. Of the remaining wagtail species, the three Afrotropical (*M. aguimp*, *M. capensis*, *M. clara*) and Malagasy (*M. flaviventris*) species exhibit slight or no geographical plumage variation and are resident [13,14].

Much of wagtail divergence is thought to have occurred during the Pleistocene, either through dispersal [17] and/or mediated by climatic events [19]. Climate changes caused by cyclical glaciation and aridification during the late Pliocene through the Pleistocene had drastic effects on many modern day species distributions [23–25]. The ancestors of modern populations would have had to seek refuge in isolated pockets of suitable habitat, potentially resulting in population differentiation due to genetic drift and natural selection in isolation. Three major refugia are thought to have periodically existed in the eastern, western, and southern edges of Eurasia since the late Miocene [24].

While the extent to which climate oscillations affected distributions is species-specific [26], both *M. alba* and *M. flava* display geographic variation in plumage types (i.e., subspecies) beyond the number of plausible refugia [24]. The "excess" of *M. alba* and *M. flava* subspecies in relation to refugia could be due to waves of isolation corresponding with past climate oscillations, which were punctuated by periods of sympatry that reinforced sexual or social character displacement over the course of multiple cycles [27]. Individual wagtail species may have responded to these changes differently, or may have been impacted by the same events [28]. Alternatively, differentiation could have occurred after dispersal out of glacial refugia following the last glacial maximum, and due to rapid local adaptation or local sexual selection across the wide distribution of these birds.

In this study, we utilize genome-wide SNPs, nuclear introns, and mtDNA to analyze speciation patterns in *Motacilla*, with complete species-level sampling and comprehensive coverage of the three most diverse Palearctic species. We (1) estimate the first time-calibrated species tree for this group; (2) demonstrate conclusively that mtDNA alone is inappropriate for phylogenetic studies of *Motacilla*; (3) investigate the agreement between genotype and phenotype, along with the proposed parallel and rapid evolution of plumage in the three most variable wagtail species; and (4) model past demographic events to reveal the temporal dynamics of effective population size and investigate how past climatic events influenced *Motacilla*.

**Materials and methods**

*Sanger data*

To resolve species-level relationships within *Motacilla*, we utilize previously published and unpublished sequences from (1) three nuclear introns (*CHD1Z*, *ODC*, *Mb*) for 42 individuals across all 12 *Motacilla* species [16,20], and (2) two mitochondrial regions (*ND2, CR*) for 103 individuals across all species, including all subspecies of *M. alba, M. flava,* and *M. citreola* (Table S1) [15–17,19]. *Dendroanthus indicus*, *Anthus pratensis*, and *Anthus trivialis* were used as outgroups [22].

*ddRADseq data*

*Sampling* - If wagtail divergence was recent or shaped by rapid ancestral radiations, then the timing between divergence events may have been too short for the emergence of phylogenetically informative mutations [29,30], potentially leading to the mito-nuclear discordance shown in previous studies. We therefore enhanced our inferential power by collecting thousands of genome-wide SNPs.

Throughout the manuscript, we follow the taxonomy outlined in previous publications [13,22]. We obtained extensive geographic sampling and near-complete taxonomic coverage across Eurasia from samples at the Burke Museum of Natural History and Culture. We augmented these specimens with samples from other natural history museums to provide complete sampling for the genus (Fig. S1, Table S1). A total of 246 birds were sampled from 11 of the 12 recognized *Motacilla* species [13,22]. As the goal of our study is to elucidate processes responsible for population divergence, our sampling focused on the widespread migratory, Eurasian wagtail species (*M. alba, M. flava, M. citreola, M.cinerea*) with multiple described subspecies. We examined museum skins and assigned individuals to subspecies using morphological criteria outlined in Alström and Mild (2003). We were unable to include *M. maderaspatensis* in our SNP sampling due to a lack of available high-quality tissue samples. For rooting phylogenetic trees, we sampled three individuals from the monotypic sister genus (*Dendroanthus*) to serve as outgroups.

*Data collection* - For detailed information on DNA extraction, library construction, and sequencing, see the Appendix. We generated SNPs using the double-digest restriction site-associated DNA sequencing (ddRADseq) protocol following methodology described in Peterson *et al.* (2012) [31]. A total of five lanes were sequenced (single-end reads: four 50 bp and one 100 bp). We then constructed a reference genome for a single individual of *M. alba* to improve the accuracy of our ddRADseq locus assembly (see Appendix).

*Data filtering* - The vast majority of RAD sequencing studies are conducted without a reference genome and assume that choosing one SNP per locus will result in an unlinked dataset.However, this many violate the requirements of downstream phylogenetic analyses. To explore the impact of this assumption, we compiled a "pseudo-unlinked" and a truly unlinked dataset by filtering out all but one locus within a 100 kb range. As missing data can affect downstream PCA results [32], phylogenetic inference [33,34], and other analyses as well [35], we further filtered datasets according to different levels of missing data (all possible combinations of 25%, 50%, and 75% missing loci and individuals). Together with our linkage filtering, we compiled 18 datasets per species group (two linkage treatments x three missing loci treatments x three missing individual treatments). We used all 18 datasets independently when conducting population structure analyses, Mantel tests, and RAxML phylogenetic trees. For all other analyses, we used a 50% threshold for both missing loci and individuals (see below).

*Phylogenetic analyses*

*Time-calibrated species tree and gene trees* - Nuclear DNA and mtDNA can support contradicting phylogenetic relationships and methods that ignore incomplete lineage sorting may fail to accurately estimate species tree relationships [36]. Because *Motacilla* lacks a species tree, we estimated the first time-calibrated species tree for this group using Sanger data (see above). We implemented molecular rate calibration in *BEAST v1.8.4 [37,38] using a published *Motacilla*-specific rate of 2.7% for *ND2* [19]. Recent simulation studies demonstrate that tree priors can have a large impact on divergence time dating when using datasets with mixed intra- and interspecies sampling [39]. Therefore, we conducted model selection on converged runs. As mtDNA and nuclear DNA may support different topologies for reasons other than incomplete lineage sorting, such as introgression and sex-biased dispersal, we conducted analyses on each data type independently. Using nuclear introns, we estimated a species tree in *BEAST. We also estimated separate nuclear and mtDNA gene trees in BEAST. See Appendix for details.

*Concatenated SNP tree* - To place subspecies in a wider phylogenetic context, we conducted ML phylogenetic analyses using concatenated ddRAD loci using RAxML v8.2 [40]. Due to their large population sizes and assumed recent divergences, wagtails contain a high level of heterozygosity with few fixed SNPs: over 99% of all variable sites contain at least one individual with a heterozygous SNP. Most methods count heterozygous sites as missing data and researchers have typically excluded these sites. We implemented the method of Lischer *et al*. (2013) [41] to generate 500 random haplotype samples from sequences with multiple heterozygous sites. We then inferred ML phylogenies using RAxML for each of these datasets. To account for potential SNP ascertainment bias, we implemented the Felsenstein correction [42]. See Appendix for details.

*SNP species tree* - To estimate a species tree, we implemented SNAPP [43]. SNAPP uses biallelic loci and requires at least one representative SNP from each species at each locus. Species assignments were based on population structure estimates (see below), and individuals with admixture were excluded to avoid model violations and branch length underestimation [44]. Because SNAPP is computationally intensive, convergence issues prevented us from using our full sampling scheme. We therefore ranked individuals by missing data and admixture proportions, and only included the top four from each population. All individual assignments were made with >95% posterior probability. Mutation rates (u, v) were both fixed at 1 and default parameters were used for the gamma prior (alpha 11.75, beta 109.73). Altering parameter values for the prior distributions on population size and tree length to better reflect wagtail population history caused convergence failure. For further information on running SNAPP, see Appendix.

*Population genetic structure*

Grouping individuals based on phenotype may be misleading, especially given that current wagtail taxonomy does not reflect the mitochondrial or nuclear trees [13,21,45]. Objective, genetic-based methods are preferable for inferring the number of genetically distinct populations and the assignment of individuals to those populations. As published mtDNA-based studies suggest *M. flava* and *M. citreola* are polyphyletic [16,17,20,21,46], we initially ran all population structure analyses on these two species combined. A similar approach was taken for the "African" clade (*M. clara, M. capensis, M. flaviventris, M. bocagii*) and the "black-and-white wagtails" (*M. alba, M. aguimp, M. samveasnae*, and *M. grandis*). Focal Eurasian wagtail species were further analyzed independently.

To *de novo* identify the optimal number of clusters in our data, we implemented two methods: the model free discriminant analyses of principal components (DAPC) in adegenet [47,48] and the model-based maximum-likelihood (ML) method, ADMIXTURE v1.3 [49]. One caveat of ADMIXTURE is that it assumes discrete ancestral or parental populations. When organisms exhibit continuous spatial population structure, recent studies tend to use PCA methods like adegenet [50]. However, adegenet has the undesirable behavior of assigning individuals to populations with unrealistically high probability. Therefore, to reduce the impact of their respective biases on downstream analyses, we employed these methods in concert to estimate *K* and individual assignment probabilities.

First, we ran ADMIXTURE with variable numbers of clusters *K*=1-10. We then plotted 10-fold cross-validation values terminated with default criteria, to choose the optimum value of *K*. Second, we ran the k-means clustering to assess groups using both AIC and BIC. We implemented DAPC to maximize differences between groups while minimizing variation within groups. To assess how many PCs to retain, we used cross-validation (*xvalDapc*) with 100 replicates and retained the number of PCs with the lowest mean squared error.

To assess genetic differentiation among phenotypes, we used DAPC to find the largest distance between subspecies defined *a priori*. For each species group, we ensured that a minimum of one individual per subspecies per locus was present. On average, this additional filtering step reduced our dataset by 40%. To determine the diagnosability of subspecies groups, we then compared our DAPC results to analyses with randomized subspecies definitions. Because rare alleles may be younger than common alleles, and may track more recent demographic or selective events [51], we explored their effect on population genetic inference by alternatively 1) pruning of all sites with minor allele frequency<10% or 2) building a matrix with only low-frequency alleles.

*Isolation by distance*

Populations in close proximity are expected to be more genetically similar than those located farther apart [52]. To explore whether our estimates are the result of the clustering of individuals with distinct allele frequencies or structure due to separation in space, we conducted Mantel tests on each species using the *mantel.randtest* function in ade4 [53] and ran these for 1 million permutations. To account for Earth's curvature, geographic distance was calculated using the Great Circle distance in sp [54].

Given that the ability of Mantel tests to detect isolation by distance (IBD) has been a recent area of debate [55,56], we also implemented the Estimated Effective Migration Surfaces (EEMS) method [57] to model the relationship between geography and genetics. EEMS allows visualization of variation in effective migration across each species' breeding distribution and the identification of corridors and barriers to gene flow by implementing a stepping-stone model over a dense grid. We used species' breeding ranges [58] and a dense grid of equally sized demes. To show that our results were independent of grid size, we ran each analysis using 250, 500, and 750 demes from three independent chains for 20 million MCMC iterations with a 10 million iteration burn-in. We first ran a series of short preliminary runs to choose parameter values that gave acceptance ratios between 20-30%. Graphs were constructed using rEEMSplots [57]. We visualized each run separately and checked convergence of MCMC runs (log posterior plots and Gewke diagnostic tests, Heidelberger and Welch test in coda [59]) before combining across runs and grid numbers to construct final consensus graphs.

*Demographic history*

To infer demographic history of each focal Eurasian species, we conducted ML parameter estimation in *dadi* [60] using a joint allele frequency spectrum and our population structure estimates. To narrow the number of possible two-population (2D) models, we first optimized one-population models and compared support for instantaneous versus exponential population size change. Instantaneous change was preferred in all populations.

We next considered six alternative 2D models which offered simplified but reasonably realistic representations of plausible demographic processes. The simplest model is of allopatric divergence in which ancestral population ($N_a$) splits into two populations ($N_1$ and $N_2$) that evolve in isolation for $T_s$ generations. We then considered a scenario of divergence followed by instantaneous population change beginning $T_g$ generations ago and resulting in populations of size ($N_{1g}$ and $N_{2g}$). Finally, we incorporated asymmetric and symmetric gene flow into each of these base models. Since cycles of population expansions and contractions have been a common feature of many birds [25], we tried incorporating additional stages of population size change into our models. Unfortunately, we were limited by the size of our SNP matrix and we were unable to optimize these increasingly complex models.

All data were polarized using outgroup SNPs and an extra parameter was included in our 2D models to account for misidentified SNPs. To ensure missing data was not affecting our results, we projected each dataset down to 60%, 70%, and 80% of individuals present at each locus. Each model was optimized in at least five independent runs and convergence was assessed by achieving similar likelihood scores and parameter estimates. The best diffusion fit was used for comparing models using AIC scores to account for variable numbers of parameters estimated in each model. Since the models are nested, we conducted an adjusted likelihood-ratio test using the Godambe Information Matrix and 100 bootstrap replicates to check whether there is support for the model preferred by AIC (complex model). To do this, we evaluated the best-fit parameters from the simpler model using the complex model. Credible intervals were calculated using the Fisher Information Matrix (*FIM_uncert*). Finally, optimized parameters were converted to real time and population size units using a generation time of one year. Mutation rates are lineage-specific and scale with divergence time, therefore we conducted conversions using a range of mutation rates: a flycatcher germline specific rate of $2.3 \times 10^{-9}$ m/s/y [61], a zebra finch lineage rate of $2.2 \times 10^{-9}$ m/s/y [62], and the mean galliform mutation rate of $1.35 \times 10^{-9}$ m/s/y [63].

**Results**

We successfully constructed a reference genome for *M. alba* which was used to assemble ddRADseq loci. A total of 219 million quality filtered reads were aligned with 5x average coverage to 90% of the zebra finch genome. Reference mapping of 442.5 million ddRADseq reads resulted in $8.2 \times 10^4$ unique loci across 246 individuals with an average coverage of 29.7 (BioProject PRJNA356768). See Table S1 for individual-level details.

For each species group, we analyzed 18 different dataset combinations consisting of either pseudo-unlinked or unlinked SNPs, and varying levels of missing data at the locus and individual level. On average, controlling for linkage reduced each dataset by 70%. This SNP pruning has potential to either remove biased SNPs or, if linkage was not a problem, to remove relevant information from the analysis. However, linkage did not alter our population structure estimates (Table S2) or Mantel tests (Table S3). Overall, filtering data based on missing data at the individual level had a larger impact than filtering data based at the locus level (Table S2). As individuals with excess missing data were removed, population structure estimators were less likely to find a consistent result than when loci were removed. We present the majority-rule analyses (50% missing data) and "pseudo-unlinked" SNPs.

*Phylogenetic analyses*

The *Motacilla* relationships inferred in our time-calibrated species tree (Fig. 1c) were confirmed by genome-wide SNPs (Fig. 1b). Relationships were, for the most part, consistent with current wagtail taxonomy, but incongruent with the mtDNA tree (Fig. 1a).

*Mitochondrial relationships* - The mtDNA tree (Fig. 1a) recovered an "African clade" consisting of four Afrotropical/Malagasy species and an "Eurasian clade", which includes the Afrotropical *M. aguimp*. Consistent with previous mtDNA studies, both *M. flava* and *M. citreola* are polyphyletic within the Eurasian clade. *M. flava* is separated into a western clade (Z) and an eastern clade, which is further divided into two sub-clades (X, Y). Only clade Y contains a monophyletic subspecies grouping of *M. f. tschutschensis*. Each *M. flava* clade has a corresponding *M. citreola* clade: *M. c. citreola* is split between clades X and Y, and clade Z includes the southern *M. c. calcarata*. There is no structure consistent with subspecies designations in *M. alba* or *M. flava* clade X or Z.

*Nuclear relationships* – Both the RAxML gene tree analysis of the concatenated SNP data (Fig. 1b) and the SNAPP species tree analysis (Fig. 1d) support the African and Eurasian clades, with the latter divided into two primary clades notable for their plumage coloration: clade A includes only "black-and-white" (melanin-based) species, whereas clade B contains only species with green/yellow (carotenoid-based) plumages. These nuclear relationships are strikingly incongruent with mtDNA in 1) placing *M. aguimp* within the "black-and-white" clade, 2) supporting the monophyly of both *M. flava* and *M. citreola*, and 3) placing *M. bocagii* within the African clade. In the SNAPP tree, the Afrotropical mainland species (*M. capensis* and *M. clara*) form a clade, whereas the insular species (*M. flaviventris* and *M. bocagii*) form another. RAxML (Fig. 1b) found high support for splits among species but no support for within-species relationships. Accordingly, within *M. alba*, *M. flava* and *M. citreola*, no subspecies are monophyletic.

*Time-calibrated species tree* - The topology of the time-calibrated species tree (Fig. 1c) is identical to the SNAPP tree (Fig. 1d); moreover, it includes *M. maderaspatensis*, for which no SNP data are available. The African and Eurasian clades split towards the end of the Pliocene (~3 mya), with early and relatively widely spaced divergences from the early Pleistocene in the former clade, and an explosive radiation within < 0.5 million years (my) during the late Pleistocene.

*Population structure*

At the species level, *de novo* population estimation methods were able to distinguish recognized species, except *M. aguimp*, which was grouped with western *M. alba* by both DAPC (Table S2) and ADMIXTURE. Narrowing our focus to phenotypically distinct subspecies, the full dataset showed genetic structure discordant with phenotype in all species complexes. Instead, eastern and western populations (*K=2*) were resolved by both ADMIXTURE and *de novo* DAPC in *M. citreola* (2 subspecies, Fig. 2) and *M. flava* (13 subspecies, Fig. 3, Table S2). Depending on which *M. alba* (9 subspecies) + *M. aguimp* dataset was analyzed in DAPC, there was support for either *K=2* or *K=3* (Table S2). ADMIXTURE consistently supported *K=2* (Fig. 4c). For all three datasets, individual assignments were consistent across methods when *K=2*. Finally, we found no difference in population structure when rare variants were either considered separately or ignored.

Ordination plots from DAPC analyses with subspecies defined *a priori* provide additional resolution of population structure in *M. flava* (Fig. 3b) and *M. alba* (Fig. 4b). In both species, the x-axis of the ordination plots show a correlation between diagnosability and longitude. In *M. flava*, all subspecies clusters overlap (Fig. S3), suggesting genetic distinctiveness is limited, especially since DAPC will maximize the little differences that are present. We note our sampling of *M. f. tschutschensis* included individuals from the easternmost portion of its range. However, *M. f. tschutschensis* is distributed from Central Eurasia through the Kamchatka Peninsula and into Western Alaska with its geographic center close to that of *M. f. macronyx* and *M. f. taivana*. Therefore, its position in the center of the DAPC ordination plot conforms to the idea that this pattern tracks longitude, and it is unclear whether these clusters correspond to subspecies distinctions or merely IBD.

The ordination plot generated by assigning *M. alba* samples to subspecies is influenced by missing data. The less stringent the threshold, the more distinct *M. a. subpersonata* becomes. This trend is found in both pseudo-unlinked and unlinked datasets, with the unlinked presenting the most extreme cases (Fig. S4). However, it is remarkable that in three out of four cases, *M. a. subpersonata* is more distinct than *M. aguimp* (Fig. S4) and all other *M. alba* subspecies (Fig. 4b). As no other subspecies displayed this behavior, we removed *M. a. subpersonata* and recompiled datasets. This resulted in *M. aguimp* samples being tightly clustered and diagnosable in all analyses (Fig. S5). The remaining *M. alba* subspecies clusters are overlapping and do not show clustering consistent with subspecies, but do show clustering on either side of the vertical axis that is consistent with our *de novo K=2* clusters. Taken together with evidence that Bayesian clustering methods may not work well with small sample sizes [64], we consider both *K=2* (without *M. a. subpersonata*) and *K=3* (*M. a subpersonata* included as own population) population histories in the following analyses.

Population structure in *M. cinerea* was not consistent across methods. Whereas ADMIXTURE supported a single panmictic population, DAPC supported an east-west split (*K=2*) consistent with the other Eurasian wagtails. These contrasting findings may be explained by simulation and other empirical studies, which have demonstrated that Bayesian clustering methods fail to detect structure when genetic divergence is very low [65,66].

*Isolation by distance*

When dealing with genetic data from evenly spaced samples from a spatially structured organism, the expected behavior of PCA is to return clusters that are related to the geographic origin of each individual sample [50]. However, spatial autocorrelation may bias interpretation of PC analyses. While Mantel tests generally support a history of IBD in all species groups (p-value<0.05), p-values vary according to missing data (Table S3).

As recent studies demonstrate that sampling design can strongly bias interpretations of Mantel tests [56], we further explored IBD using EEMS, a program that can distinguish whether support for IBD is the result of either geographically distant and differentiated populations or a continuous cline in genetic differentiation. Strong linear relationships between predicted and observed genetic dissimilarities confirm that the EEMS model fit our data [57]. To assess support for a true barrier, we examined plots of dissimilarity between pairs of sampled demes for non-linearity and resulting effective migration map for a singular uniform barrier. EEMS strongly supports the existence of a barrier between eastern and western *M. citreola* (Fig. 2), consistent with the findings of population structure estimators.

In *M. cinerea*, EEMS finds no support for a barrier, consistent with population structure estimates from ADMIXTURE (*K=1*) but not DAPC (*K=2*). We consider *M. cinerea* a single panmictic population with a strong signal of IBD in downstream analyses. This is consistent with a previously described pattern of shallow clinal plumage variation seen across much of Eurasia [13]. We were unable to include samples from two of the three subspecies (small populations restricted to either the Azores and Madeira Islands) and can only conclude that there is support for a single panmictic mainland Eurasian *M. cinerea*.

We found conflicting support for IBD in *M. flava*. Despite a linear trend in predicted and observed genetic dissimilarities, supporting IBD, there is also strong support for the existence of two discontinuous barriers (Fig. 3c). We interpreted this to mean that while IBD exists across the whole *M. flava* range, there are areas where gene flow is not significant. For assignment of subspecies into these populations, see Figure 1d and Figure 3. Since these results differ from those of *de novo* adegenet and ADMIXTURE, we tested whether DAPC can distinguish between these groups when given individual assignments. First, we fixed *K=3* and ran *de novo* DAPC to see if we could find the same populations resolved by EEMS. Second, we defined EEMS populations and ran DAPC. As both methods overwhelmingly supported the three populations found by EEMS, we used these definitions in all remaining analyses.

*M. alba* also showed a linear trend and a patchy effective migration map, lending support for IBD. However, a closer look at posterior mean of effective migration rates demonstrates that some areas have a markedly higher rate of migration than average (Fig. 4c). EEMS largely supports the east-west division of *M. alba* and the distinctness of *subpersonata*. However, an inverted Y-shaped migration surface splits eastern *M. alba* into northern (congruent with *M. a. lugens* and *M. a. ocularis*) and southern (congruent with *M. a. leucopsis* and *M. a. alboides*) populations (Fig. 4c). To explore the north-south division further, we implemented DAPC, but it was unable to distinguish between these populations. These results may stem from low sample sizes, as EEMS is more robust to biased sampling than PCA methods [57].

*Demographic history*

To narrow the number of possible 2D models, we first ran 1D models on each population separately. *M. citreola*, *M. flava*, and *M. alba* were all modeled using 2D models. (We were unable to optimize 3D *dadi* models due to a sparse site frequency spectrum, and therefore, conducted pairwise 2D model comparisons for *M. flava* [per comm. R. Gutenkunst].) All six models were optimized and at least five independent runs converged on similar parameter values. Within species, missing data did not affect demographic model selection and produced overlapping 95% credible interval of parameter estimates (Fig. S6). Therefore, we chose to present the 80% threshold in both paper figures and 2D site frequency spectra (Fig. S7-S9).

All species support the *split + growth + symmetric migration* model (Fig. 5a, Table S5), where population size increases following divergence (Table S4). All converted time estimates lie within the Pleistocene (0.14-1.8 million years ago; Fig. 5b). Time estimates largely overlap across *M. flava* and *M. citreola* populations, whereas time estimates for *M. alba* are more recent. Real time estimates are sensitive to the choice of multipliers, such as generation time and mutation rate. Therefore, we were primarily interested in relative population size and relative time estimates. However, all three avian mutation rates recover time estimates within the Pleistocene (Fig. S6).

Finally, *M. cinerea* consists of a single panmictic population supporting a two-epoch model, where at time *T* the ancestral population underwent instantaneous population growth.

**Discussion**

Modern phylogeographic studies assume that the population divergence history of widespread and variable species can be disentangled with sufficient numbers of variable, neutrally evolving SNPs [67,68]. We collected genome-wide SNPs to compare populations at different stages of the speciation continuum. Our data resolved species-level relationships with high support, suggesting that some populations are structured along an east-west axis, but failed to distinguish between recently diverged, phenotypically distinct taxa that are usually treated as subspecies. Recent differentiation, ongoing gene flow, and demographic processes may have obscured patterns of differentiation at neutral loci (cf. [69,70]).

*Phylogeny*

We provide evidence that the currently recognized wagtail species are monophyletic and resolve the relationships among them. Within the "African" clade, we find a sister relationship between mainland and island species. Within the latter clade, *M. bocagii* is endemic to São Tomé, an island off the west coast of Africa, while *M. flaviventris* is endemic to Madagascar. Originally thought to belong to the sylvioid superfamily, *M. bocagii* bears little resemblance to other *Motacilla* species in both plumage, structural morphology, habitat, and behavior [22]. Our SNP data confirms its placement within *Motacilla*.

Species belonging to the "black-and-white" clade have mostly allopatric distributions, together covering nearly all of Europe, Asia, and Africa. While our *de novo* population structure analyses are unable to differentiate between *M. aguimp* and *M. alba*, *M. aguimp* is identifiable using *a priori* assignment. Moreover, our species tree analyses (Fig. 1d & Fig S2), which account for stochastic processes such as incomplete lineage sorting, found *M. aguimp* sister to *M. alba* with high support, rather than embedded within *M. alba*. The distinctness of *M. aguimp* is further substantiated by its multiple differences (size, plumage, song, call) from all subspecies of *M. alba* [13].

We were unable to confirm the exact placement of species sister to the clade containing *M. alba* and *M. aguimp* due to the lack of *M. maderaspatensis*, a large-bodied resident of the Indian continent, in our SNP dataset. However, we confirmed monophyly of two of its members, *M. grandis* and *M. samveasnae*, which had been insufficiently sampled in previous studies. Endemic to Japan, *M. grandis* is a monotypic, short-distance migrant that is easily distinguishable from its partially sympatric relative *M. alba* by its plumage, larger body size, song and call [13]. Despite overlapping distributions and documented hybridization, no prior population genetic analyses of *M. alba* have included more than a single sample of *M. grandis* [19,71] and many have nested *M. grandis* within *M. alba* [13,19]. Using four newly sampled *M. grandis* individuals, our results show that *M. grandis* is distinct from *M. alba*. The close relative, *M. samveasnae*, has a comparatively minute distribution and is endemic to the Lower Mekong basin of Cambodia and Laos. In appearance, it closely resembles the Afrotropical *M. aguimp* but our increased sampling confirms its distinctness.

Previous mtDNA-based studies on *M. alba* delineated three or four populations [19,71]. The only population consistent with our SNP results was the rare Moroccan endemic *M. a. subpersonata*, which was found to be divergent in both studies. Li *et al*. (2016) included denser sampling of the southeastern part of the *M. alba* range, but this sampling difference is unlikely to explain the incongruent population structure and instead reflects inherent differences between mitochondrial and nuclear data, such as mutational rate and inheritance patterns. Intra-specific relationships remain unresolved and divergence time estimates support a scenario of recent, rapid radiation.

Within the yellow plumage clade, this same trend holds true for both *M. flava* and *M. citreola*. *M. flava* and *M. citreola* are monophyletic and sisters, contrary to mtDNA findings and accordingly does not support a previous proposal to split *M. citreola* into two species [18].

*Plumage divergence*

Discordance between phenotype and genotype, along with our divergence time estimates, suggest that wagtail plumage evolution has been very recent and rapid. This is particularly true for the Palearctic *M. flava* and *M. alba* complexes, which display exceptionally high plumage differentiation during the second half of the Pleistocene (as also suggested by [17] for *M. alba*). This has led to some remarkable cases of parallel evolution, e.g. in the widely disjunct *M. aguimp* vs. *M. samveasnae*; *M. f. thunbergi* vs. *M. f. macronyx*; and *M. f. flava* vs. *M. f. tschutschensis* (cf. Fig. 1). Moreover, the widely allopatric *M. f. flavissima*, *M. f. lutea* and *M. f. taivana* differ from the other *M. flava* subspecies by their bright yellow and green head patterns, and in the *M. alba* complex certain head pattern and upper part coloration traits display a mosaic geographical pattern, indicating parallel evolution.

Recent studies have demonstrated that strong selection can occur at few genes, and that plumage differences can evolve rapidly without corresponding divergence in the rest of the genome [72,73]. For example, within a Eurasian crow (*Corvus*) species complex, there is evidence of assortative mating based on plumage despite no genetic differentiation across most of the genome. Instead, there appears to be simple genetic control of phenotype, but divergence is occurring at different genes in different populations, thus implicating a more complex multi-genic pathway than previously thought [74]. While our SNP data was sampled from across the genome, it only represents 0.05% of the entire 1.1 Gb wagtail genome. If few genes are responsible for plumage differences and populations experienced recent divergence and selection, then it is not surprising that our data fails to find a genetic signal congruent with phenotype. Strong selection on plumage loci, mediated through assortative mating and selection against intermediate plumage phenotypes (hybrids), might explain how different plumage traits can be maintained in the face of gene flow in hybrid zones. Hybridization is prevalent, in variously wide hybrid zones, among parapatrically distributed subspecies of *M. flava*, *M. citreola* and *M. alba*, and interspecific hybridizations, e.g. between *M. flava* and *M. citreola*, are not infrequent (reviewed in [12]). However, evidence for decreased hybrid fitness is lacking.

Strong integrity of the *M. f. thunbergi* phenotype, but large-scale introgression of other loci from *M. f. tschutschensis* might perhaps explain the confusing pattern in the continuously distributed *M. f. thunbergi* (cf. Fig. 3), which displays a particularly strong genotype vs. phenotype discordance. Such a pattern has been found in the *Corvus corone* complex in western Europe, where adjacent populations of the phenotypically different *C. c. corone* and *C. c. cornix* are genetically more similar than some phenotypically similar populations of the former [72]. Alternatively, the *M. f. thunbergi* phenotype might have 'invaded' the *M. f. tschutschensis* genome through a selective sweep. Introgression of specific morphological traits have been noted in e.g. *Heliconius* butterflies [75] and domestic chicken [76]. However, neither of these explanations seems fully compatible with the genetic differentiation and gene flow restrictions between western and eastern *M. f. thunbergi* first detected by mtDNA [16] and confirmed by SNP data in the present study.

In general, *Motacilla* species vary considerably more in plumage traits closely linked to signalling functions than to ecological (e.g. beak, tarsus, or wing length) traits (cf.[77]), indicating that sexual selection might have played an important role in phenotypic divergence. Furthermore, the Eurasian clade displays variation in sexual dimorphism, size, seasonal plumage variation, and age-related differences, whereas the African clade does not[13], indicating greater potential for sexual selection in the former group. However, field experiments are necessary to confirm the role of plumage in mate preference. Sexually selected plumage traits have potential to diverge rapidly among allopatric, ecologically equivalent populations [78], and have been identified as the dominant signals mediating mate-choice and intrasexual aggression in some birds [77]. In contrast, the aberrant structure and plumage of *M. bocagii* (cf. Fig. 1) has been suggested to be the result of natural selection following colonization of a novel habitat [21].

*Demographic history*

Many bird species have experienced range shifts in the Palearctic since the late Miocene [24,79,80]. However, evidence of avian divergence during the Pleistocene varies [25,81,82]. Here, we explicitly modelled patterns of population divergence, population size changes over time, and patterns of gene flow, placing Eurasian wagtail diversification occurring during the Pleistocene. Previous divergence estimates, based solely on mtDNA gene tree calibration, are much earlier [17,22], but are likely biased by multiple missing taxa.

If the pattern of intraspecific divergence in the *M. alba* and *M. flava* species complexes was caused by fragmentation of breeding ranges and prolonged periods of divergence in isolation, then we would expect to see corresponding genetic differentiation among subspecies and demographic models supporting deeper divergence times. However, if multiple interglacials allowed populations to repeatedly meet and mix, differentiation evolved in isolation could have eroded. Under this scenario, it also possible that differential selection on plumage controlled by few genes could lead to the plumage diversity we see today. Alternatively, if subspecies are the result of recent and rapid selection following the end of the last glacial maxima, we would expect to see population structure incongruent with phenotype and a recent history of population expansion in demographic models.

While neither scenario is consistent with our *M. flava* or *M. citreola* results, we do find support for the latter in *M. alba* which bears the mark of the most recent climatic events. In *M. alba*, the 95% credible interval of population division (Fig. 5b) coincides with the start of a cooling period (~75,000 ybp; [83], which lasted until the last glacial maximum (LGM: ~22,000 ybp; [84]. Following the LGM, ice sheets receded and our estimates find a corresponding population increase during that time. These findings are consistent with past work on the *M. alba* species complex which also found evidence for Pleistocene population expansion and relatively recent introgression at the edge of their range [19,71]. Furthermore, the upper 95% credible interval divergence time estimates (Fig. 1c) predate those of our population divergence time estimates (Fig. 5b) and are therefore consistent.

Coincident patterns of barriers to gene flow in *M. flava* and *M. citreola* suggest that major eastern and western refugia existed. The western barrier coincides with a vast area of lowland (Turgai depression) that periodically became a huge body of water, potentially acting as a west-east barrier during interglacial periods [85]. Therefore, western and eastern populations may have been separated during both glacial and interglacial periods. Shorter periods of separation in minor refugia may have then led to plumage differentiation under strong sexual selection, but without more large-scale genetic divergence. Further separation of *M. flava* into northern and southern populations in the east is likely a result of East Asia being much less affected by glaciation. Indeed, other birds show similar divergence patterns as wagtails [6,86,87], indicating a common cause. For a more detailed reconstruction of complex demographic events, increased genomic sampling is needed [88].

While still supporting a history of Pleistocene diversification, both *M. citreola* and *M. flava* divergence time estimates (Fig. 1c) postdate those of demographic models. This inconsistency could be due to two factors. First, the *M. alba* specific mtDNA rate used to calibrate our species tree was derived from the generic avian *cyt-b* rate. It is well established that molecular rates vary across lineages and this rate may be inaccurate. Second, gene flow leads to underestimation of divergence times in time-calibrated phylogenies [44]. Hybridization between *flava* and *citreola* has been documented numerous times (reviewed in [13]), predominantly on the expanding western edge of the *citreola* range, and may explain why divergence dates are more recent than our estimates population split time.

Once thought to be a panacea for resolving relationships between rapidly evolving lineages, genome-wide SNPs were unable to distinguish between phenotypically-distinct wagtail populations. It is clear that subspecies should not be treated as evolutionary units. Future studies should harness the power of whole-genome re-sequencing and gene expression studies, as changes in gene expression often underlie changes in phenotypic differentiation [89].

**Data Accessibility**

- Raw, demultiplexed ddRAD reads: NBCI SRA under PRJNA356768
- SNP datasets and input files for analyses: Dryad repository doi:10.5061/dryad.008bq

**Acknowledgements**

We thank the museums and those who sent us samples: U. Johansson, Naturhistoriske Riksmusect; B. Schmidt, USNM; I. Nishiumi, NMSM; B. Marks, FMNH; Mark Robbins, KU; Jack Withrow, UAM; P. Sweet, AMNH; K. Zyskowski, YPM; M. Westberg, Bell Museum; G. Voelker, TCWC; S. Birks, UWBM. We thank M. Melo for sharing tissues and D. Shizuka facilitating contacts in Japan.

For helpful discussion and comments, we thank S. Rohwer, J. Klicka, N. Sly, S. Billerman, E. Linck, C. Battey, and the Leache Lab. For assistance with sequencing and analysis, we thank R. Gutenkunst, N. Bouzid, D. Petkova, and J. Grummer.

R.B.H. was supported by the NSF-GRFP. Funding came from the AMNH Chapman Grant, Sigma Xi GIAR, NSF DDIG DEB-1501131, and the Burke Museum of Natural History and Culture. We used the UC Berkeley Vincent J. Coates Genomics Sequencing Laboratory, supported by NIH S10 Instrumentation Grants S10RR029668 and S10RR027303, and the advanced computational, storage, and networking infrastructure provided by the University of Washington Hyak supercomputer system. P.A. was supported by the Swedish Research Council, Jornvall Foundation, and the Constantines.

Figure 1. (a) BEAST mitochondrial gene tree, with bars indicating paraphyletic groupings of *M. flava* and *M. citreola*: X = northeastern, Y = southeastern, and Z = northern. (b) RAxML consensus tree of 500 random haplotype datasets; in both, nodes of monophyletic, single-species clades are collapsed for ease of viewing. Numbers in parentheses indicate sample size and bars indicate the two Eurasian-breeding clades: A = "black-and-white" plumage and B = "yellow" plumage clade. (c) *BEAST tree inferred from mtDNA and nuclear introns, calibrated using a 2.7% *ND2* rate. Node bars indicate the 95% HPD of height. (d) SNAPP maximum clade credibility species tree (1467 biallelic SNPs with 8.5% missing data) rooted with *Dendronanthus* (not shown). In all trees, posterior probability is indicated at the nodes, where a circle denotes PP≥0.95. In all figures, paintings by Bill Zetterström and Ren Hathway.

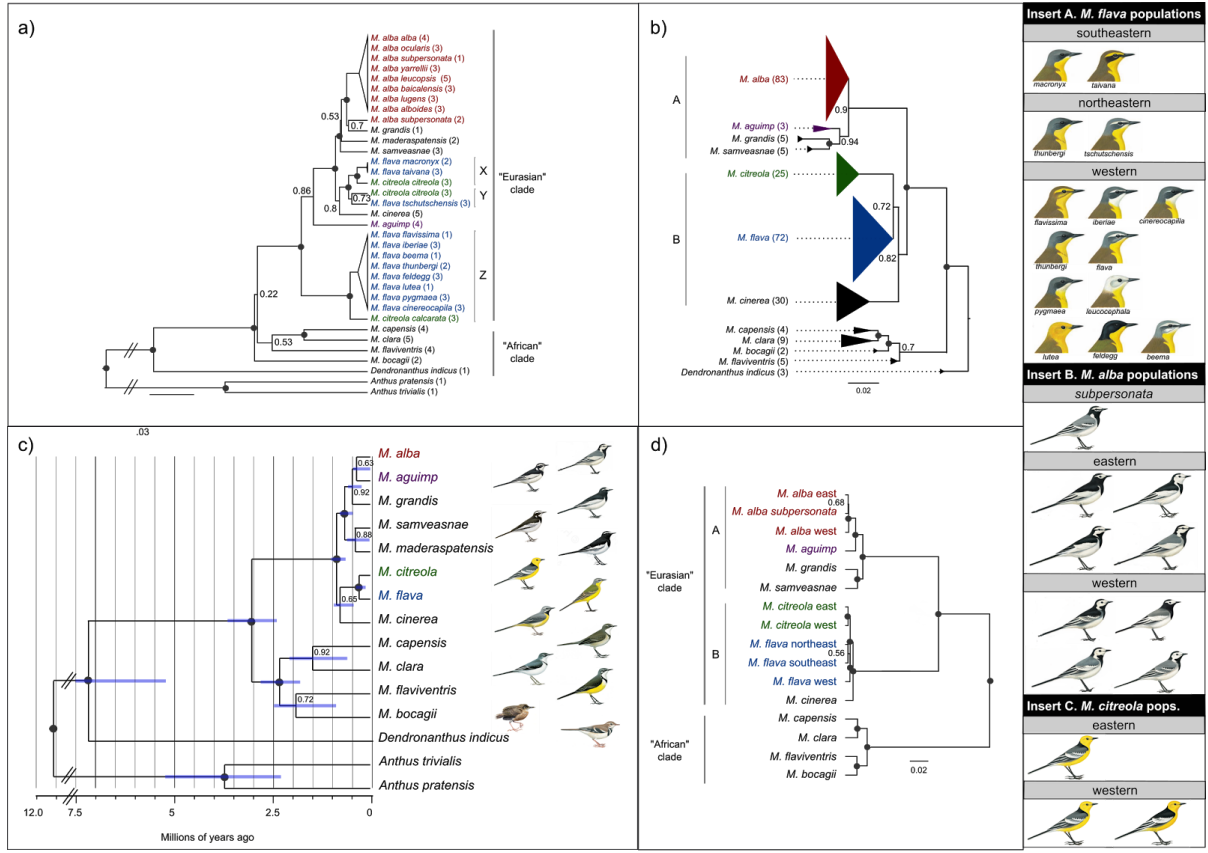

Figure 2. Posterior probabilities of effective migration rates of *M. citreola* estimated by EEMS. Birds sampled in India belong to *M. c. calcarata*, all others to *M. c. citreola*. In all figures, pie charts are located at sampling sites and denote the posterior probability of ADMIXTURE assignments (here, *K*=2).

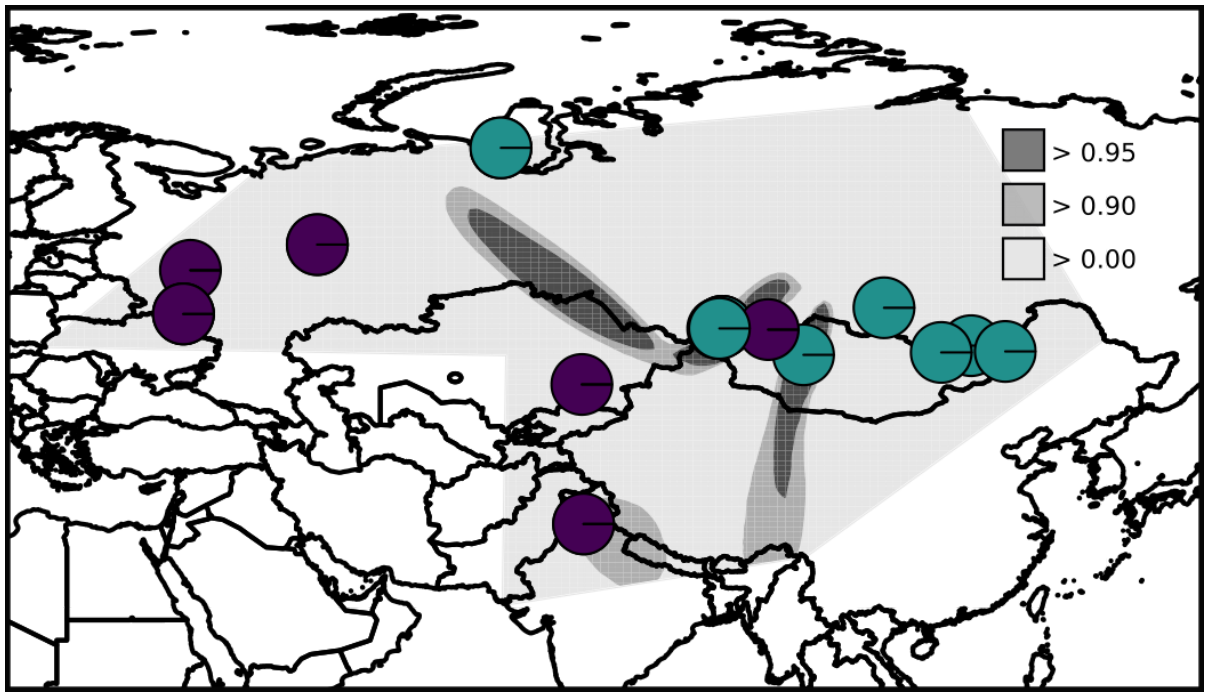

Figure 3. (a) *M. flava* breeding distribution and sampling localities. (b) DAPC plot of genetic clustering by subspecies (c) Posterior probability of effective migration rates estimated by EEMS show two barriers and *K*=3.

a)
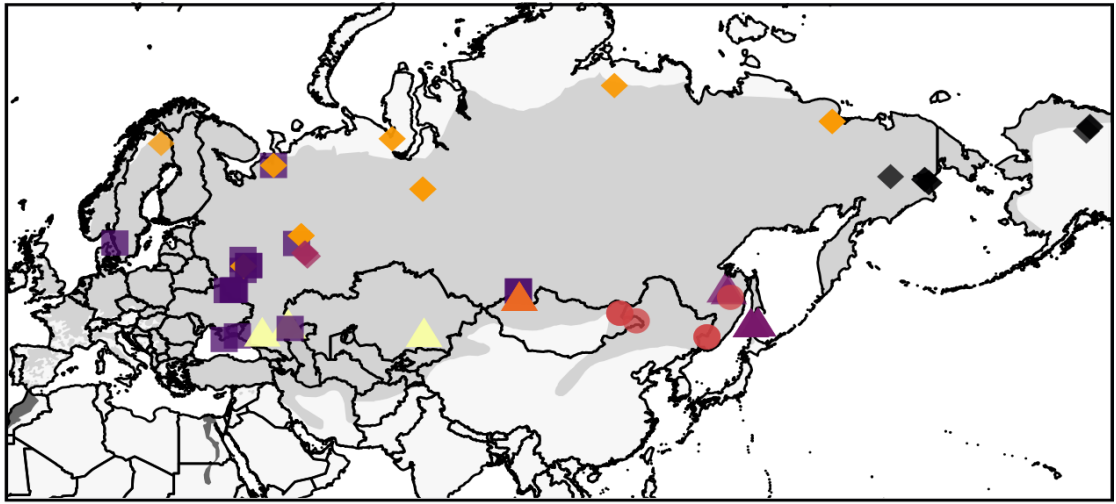

b)
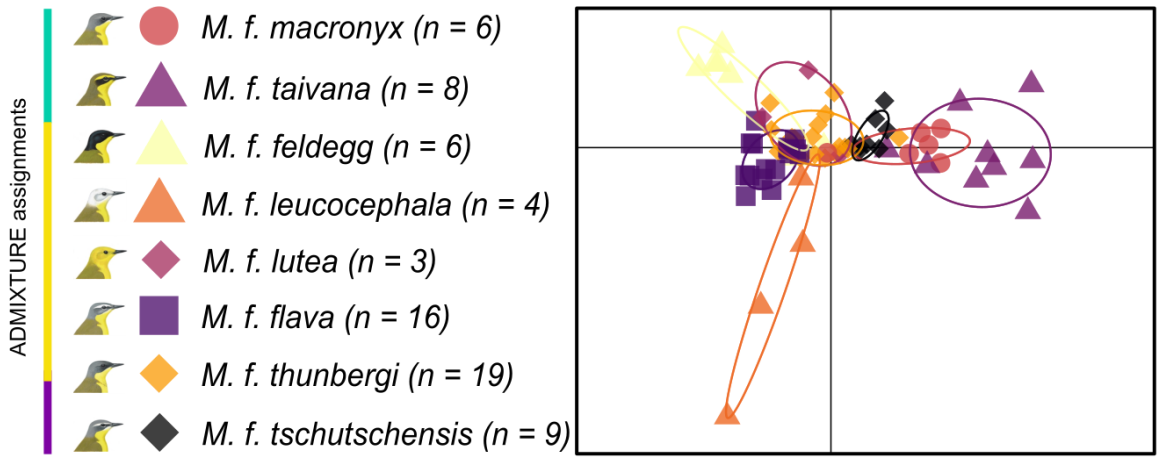

c)
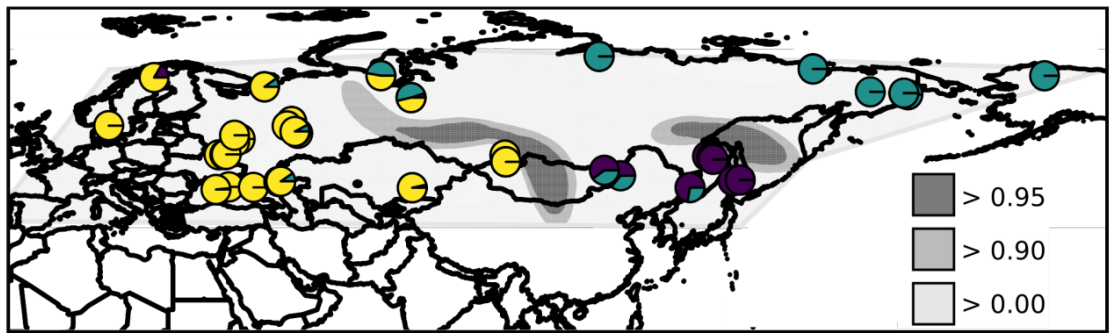

Figure 4. (a) *M. alba* distribution and sampling localities. (b) DAPC plot of genetic clustering by subspecies. (c) Posterior probabilities of estimated effective migration rates show 2-3 barriers and *K*=2.

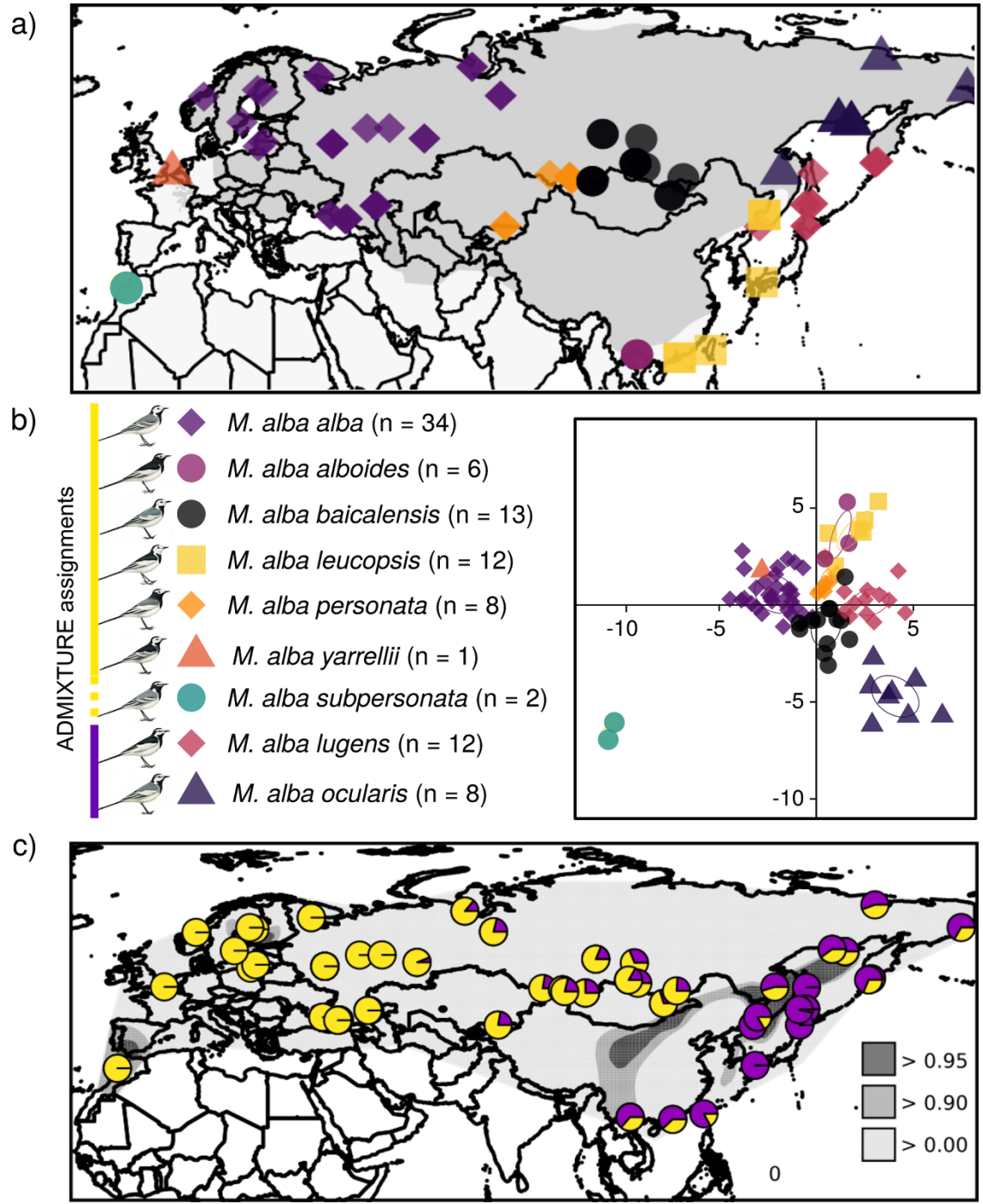

Figure 5. (a) Schematic representation of the best-fit demographic model, *split + growth + symmetric migration*. (b) Comparison of 2D *dadi* time estimates.

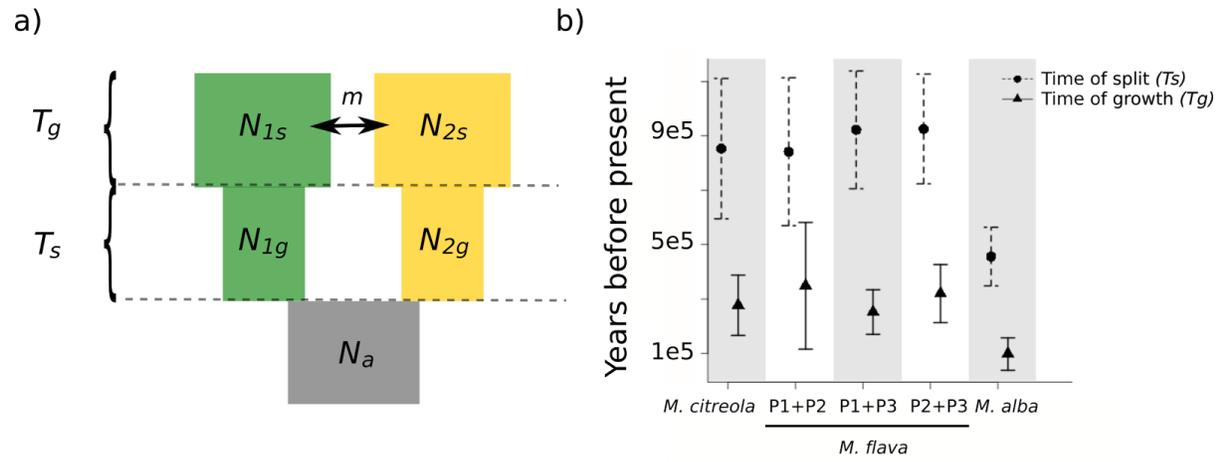